\documentclass{optica-article}

\journal{opticajournal} 

\articletype{Research Article}
\usepackage{lineno}
\usepackage{graphicx}  
\usepackage{hyperref}  
\usepackage{float}     
\usepackage{lineno}
\usepackage{tikz}
\usepackage{caption}
\usepackage{booktabs}
\usepackage{amsmath}

\usepackage{amssymb}
\usepackage{siunitx}
\usepackage{placeins}
\usepackage[percent]{overpic}
\usepackage{multirow}
\usepackage[utf8]{inputenc}
\usepackage[T1]{fontenc}

\DeclareUnicodeCharacter{202F}{\,}
\DeclareUnicodeCharacter{2009}{\,}
\DeclareUnicodeCharacter{03C3}{\ensuremath{\sigma}}
\DeclareUnicodeCharacter{221E}{\ensuremath{\infty}}

\begin{document}

\title{Deep Learning for Optical Misalignment Diagnostics in Multi-Lens Imaging Systems}

\author{Tomer Slor,\authormark{1,\dag,*} Dean Oren,\authormark{1,\dag} Shira Baneth,\authormark{1} 
        Tom Coen,\authormark{2} Haim Suchowski\authormark{2}}

\address{\authormark{1}School of Computer Science and AI, Faculty of Exact Sciences, 
                   Tel Aviv University, Tel Aviv 69978, Israel\\
         \authormark{2}School of Physics and Astronomy, Faculty of Exact Sciences, 
                   Tel Aviv University, Tel Aviv 69978, Israel\\
         \authormark{\dag}These authors contributed equally to this work.}

\email{\authormark{*}tomerslor@mail.tau.ac.il}

\begin{abstract}
In the rapidly evolving field of optical engineering, precise alignment of multi-lens imaging systems is critical yet challenging, as even minor misalignments can significantly degrade performance. Traditional alignment methods rely on specialized equipment and are time-consuming processes, highlighting the need for automated and scalable solutions.
We present two complementary deep learning-based inverse-design methods for diagnosing misalignments in multi-element lens systems using only optical measurements. First, we use ray-traced spot diagrams to predict five-degree-of-freedom (5-DOF) errors in a 6-lens photographic prime, achieving a mean absolute error of \SI{0.031}{\milli\meter} in lateral translation and 0.011$^\circ$ in tilt. We also introduce a physics-based simulation pipeline that utilizes grayscale synthetic camera images, enabling a deep learning model to estimate 4-DOF, decenter and tilt errors in both two- and six-lens multi-lens systems. These results show the potential to reshape manufacturing and quality control in precision imaging.
\end{abstract}

\section{Introduction}
Precise optical alignment is critical for ensuring high performance in multi-element imaging systems. Slight translational or angular misalignments between lens elements can introduce aberrations that severely degrade image quality. Conventional methods, including Hartmann tests, interferometry, and star-target diagnostics, typically demand time-consuming manual adjustments, dedicated equipment, or significant expertise, limiting their use in high-volume manufacturing or real-time field calibration \cite{malacara2007,smith2008MOE,primot2000extended,crawford2008,wyant2010,luna1999}.

As demand grows for high-performance optics in aerospace, medical, and consumer devices, there is an urgent need for automated, efficient, and reliable alignment solutions. Data-driven approaches have begun to emerge, as early as 1993, NASA demonstrated a neural network guiding a laser-beam alignment system \cite{Wang1993NeuralAlignment}. More recently, deep learning (DL) has shown promise in a range of optical metrology tasks, including surface defect inspection, wavefront sensing, and even automatic lens design optimization \cite{Zuo2022DeepLearningOpticalMetrology,Cote2021,Hegde2019}.
Recent efforts have extended DL into production workflows, enabling self-optimizing and genetic-algorithm-aided lens assembly \cite{Holters2016,Min2024}, nodal-aberration-based tolerance optimization \cite{Gu2020}, and active alignment of camera modules in factory environments \cite{Liu2024ActiveAlignment}.

Learning-based misalignment diagnosis now spans reflective telescopes \cite{Jia2021TelNet,Baslar2024MisalignmentNN,Wu2022MLAlignment}, digital-twin survey optics \cite{Rothe2023SkySurvey}, tolerance-aware design frameworks \cite{Gungor2025ToleranceAware}, and automated optical adjustment via simulated through-focus imagery \cite{Hashimoto:24}. Yet most prior studies focus on single-element perturbations, assume near-symmetry, or rely on interferometric or point-spread-function inputs that are  time-consuming to acquire.
Moreover, most evaluations focus on low-dimensional fault spaces, such as two-element assemblies or rotational-only or translational-only misalignments, limiting confidence in their generalization to complex, asymmetric multi-lens systems.

In this work, we address these gaps by introducing a deep learning-based framework capable of diagnosing full five-degree-of-freedom misalignments, specifically decenter and tilt deviations along orthogonal axes, for each lens element in a multi-component system. Our approach requires only standard spot diagram or incoherent image measurements that can be acquired with basic optical test setups, without any specialized wavefront sensors or interferometry. We develop two complementary models: one operating on ray-traced spot diagram data and another on raw incoherent images. Both models are trained on large-scale synthetic datasets spanning a wide range of misalignment scenarios. We demonstrate that the models achieve high accuracy in multi-lens fault diagnosis, generalize well to complex optical assemblies, and do not require symmetry assumptions, prior calibration, or physical modifications to the test setup. This capability enables scalable, real-time fault identification and paves the way toward intelligent automated lens alignment in manufacturing.

\begin{figure}[htbp]
\centering
\begin{overpic}[width=\linewidth]{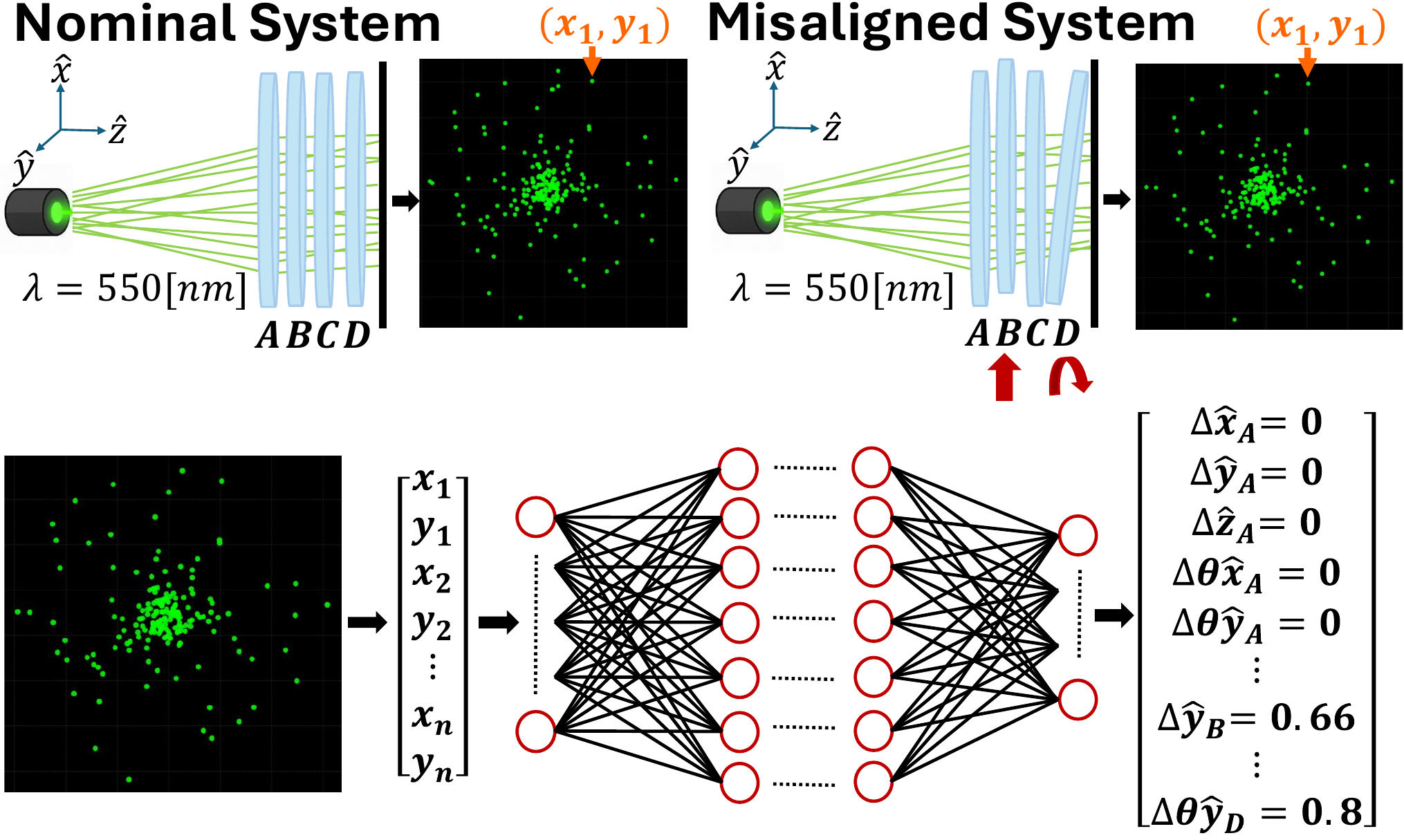}
    \put(50,32){\small \textbf{(a)}}  
    \put(50,-1){\small \textbf{(b)}}
\end{overpic}
\caption{\textbf{Overview of the misalignment diagnosis framework.} (a) Comparison between spot diagrams from a nominally aligned system and a misaligned system, where small perturbations in lens positions or orientations produce measurable shifts in the ray intersection pattern at the image plane. (b) Schematic of the learning pipeline for Method~1: the spot diagram (collection of $(x_i,y_i)$ ray hit coordinates) is flattened into a feature vector and fed into a fully connected neural network, which regresses the 5-DOF misalignment parameters (decenter $\Delta x,\Delta y,\Delta z$ and tilt about $x,y$ axes) for each optical element.}
\label{fig:overview}
\end{figure}

\section{Methods}
We developed two complementary deep learning pipelines for estimating element-wise misalignments in multi-lens optical systems. Both approaches aim to infer per-element alignment deviations (translations and tilts) from external optical data alone, without requiring internal access or interferometric measurements. Method~1 uses simulated spot diagrams from ray-tracing and a fully connected neural network regressor, while Method~2 uses through-focus intensity images and a convolutional neural network (CNN) to extract misalignment cues. Below, we describe the data generation process, model architectures, training procedures, and evaluation metrics for each approach.

\subsection{Spot-Diagram-Based Misalignment Prediction (Method 1)}

\textbf{System and data generation.} Figure~\ref{fig:overview} illustrates an overview of the spot-diagram-based method. We simulate a 6-lens photographic prime lens consisting of two cemented doublets and two singlets, based on the optical prescription in U.S. Patent US02194413-1  (see Section~\ref{app:spot}). Ray tracing is performed using a custom pipeline built on the open-source \texttt{rayopt} library~\cite{RayOpt}. Each lens element is randomly perturbed with five degrees of freedom: lateral translations $\Delta x, \Delta y$ (uniform in $[-1,1]$~mm), axial shifts $\Delta z$ (uniform in $[-1,1]$~mm where feasible and otherwise constrained to prevent element overlap), and tilts about the $x$ and $y$ axes (uniform in $[-1^\circ,1^\circ]$). No rotation is applied about the optical axis ($\theta_z$), as the lenses are rotationally symmetric. Misalignments are applied independently for each element.

The system is illuminated at two monochromatic wavelengths, 450~nm and 550~nm. Rays are traced from multiple field points to two detector planes located 16~mm and 26~mm behind the last optical surface. Each sample consists of 400 rays traced from a 2$\times$2 grid of field positions. Rays that fail to reach the screen are filtered out via ray-ID–based mask. In total, 500,000 misaligned samples are generated. The dataset is split into 80\% training and 20\% test subsets.

For each sample (i.e., misalignment instance), four spot diagrams are generated, one for each combination of the two wavelengths and two screen positions. Each spot diagram contains the $(x, y)$ hit positions of 400 rays, resulting in a total input of shape $4 \times 400 \times 2$, which is flattened to a 3200-dimensional vector.  Inputs are normalized per-ray by subtracting the mean and dividing by the standard deviation across the training set. Output misalignments are likewise normalized.

\textbf{Network architecture and training.} The model is a fully connected neural network with five hidden layers of 2048 neurons each, ReLU activations, and residual skip connections every two layers. The output is a 20-dimensional vector encoding 5-DOF misalignment values for each of the four optical elements. Training is performed using the AdamW optimizer with a learning rate of $10^{-5}$, batch size 250, and weight decay of 0.01. A Reduce-on-Plateau scheduler (decay = 0.1, patience = 10) is used alongside the optimizer which automatically reduces the learning rate by a factor of 0.1 whenever the validation loss plateaus, stabilizing convergence. The network is trained for 500 epochs on the training set using mean squared error (MSE) loss between the predicted and normalized misalignment vectors.

\subsection{Image-Based Misalignment Prediction (Method 2)}

\textbf{System and data generation.} This method uses image data generated via high-fidelity physical simulation in the 3DOptix platform \cite{3doptix2025}, which accurately models irradiance at real image planes. We consider two lens assemblies: a simple 2-lens system  (see Section~\ref{app:image2}) and a 6-lens photographic prime based on U.S. Patent US02194413-1 (see Section~\ref{app:image6}). In each case, all lens elements are independently perturbed with four degrees of freedom: lateral translations $\Delta x, \Delta y$ and tilts about the $x$ and $y$ axes. No axial shifts or $z$-axis rotations are applied.

For the 2-lens system, we generate 29,000 simulations with perturbations drawn uniformly from $\Delta x, \Delta y \in [-2, 2]$~mm and $\Delta \theta_x, \Delta \theta_y \in [-3^\circ, 3^\circ]$. Rays from an incoherent source equipped with a 1951 USAF resolution mask are propagated through the optical system to a detector placed 106~mm behind the last surface. Each simulation uses $3.3 \times 10^6$ rays at a wavelength of 465~nm. The resulting irradiance image is $1000 \times 1000$ pixels and is cropped to the minimal rectangle containing all pixels with intensity above 0.05 W/$\text{cm}^2$ per pixel. The crop coordinates $(x, y)$ and pre-crop dimensions  (width, height) are stored as metadata. Inputs are normalized and clipped.

For the 6-lens system, we simulate 95,000 perturbed configurations with $\Delta x, \Delta y \in [-0.25, 0.25]$\, mm and $\Delta \theta_x, \Delta \theta_y \in [-3^\circ, 3^\circ]$. Here we use the same incoherent source equipped with a 1951 USAF resolution mask as in the 2-lens setup and five detectors are placed at 67.5~mm, 72.5~mm, 77.5~mm, 82.5~mm, and 87.5~mm behind the last optical element. Each detector image is cropped using intensity thresholds of 10, 10, 10, 9, and 5 W/$\text{cm}^2$ for the respective detectors. The crop coordinates $(x, y)$ and pre-crop dimensions  (width, height) for each image are stored as metadata. Inputs are normalized and clipped.

\textbf{Network architecture and training.} Each image is processed by a separate ResNet18 encoder producing a 512-dimensional embedding. The associated crop metadata is processed by a three-layer MLP (dimensions [4,16], [16,32], [32,64]). These outputs are concatenated and passed through a final MLP and fully connected layer to produce the predicted deviation vector. The final output size is 8 for the 2-lens system and 16 for the 6-lens system, corresponding to four degrees of freedom per optical element.

Training is performed using the AdamW optimizer with a learning rate of 0.01, weight decay of 0.001, and batch size of 32. Training begins with an UntunedLinearWarmup phase to ramp the learning rate smoothly from $5\times10^{-6}$, after which it employs the same Reduce-on-Plateau scheduler (decay = 0.1, patience = 10) as used in the spot-diagram method. The model is trained using mean squared error (MSE) loss on normalized outputs, for 260 epochs. No data augmentation or early stopping is used.

\section{Results and Discussion}
\begin{figure}[htbp]
  \centering

  \begin{flushleft}
  \begin{overpic}[width=0.58\linewidth]{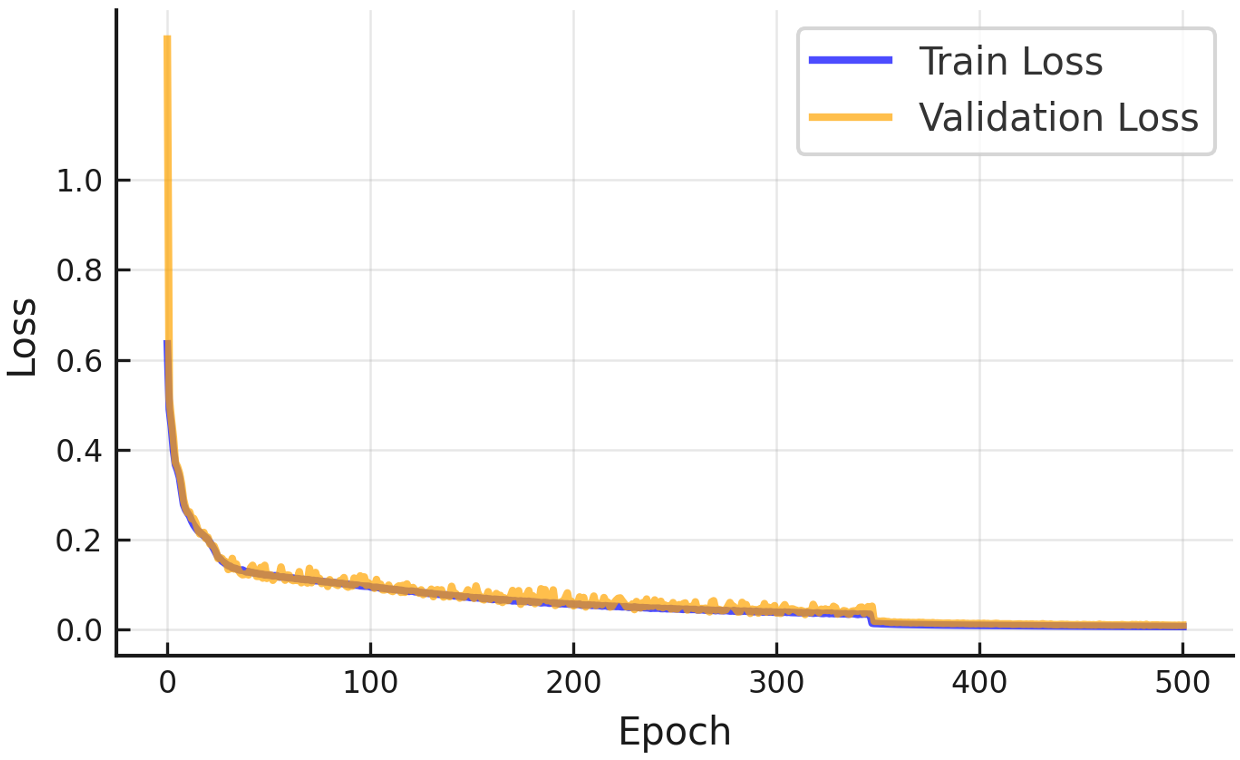}
    \put(100,8){ 
      \includegraphics[width=0.42\linewidth]{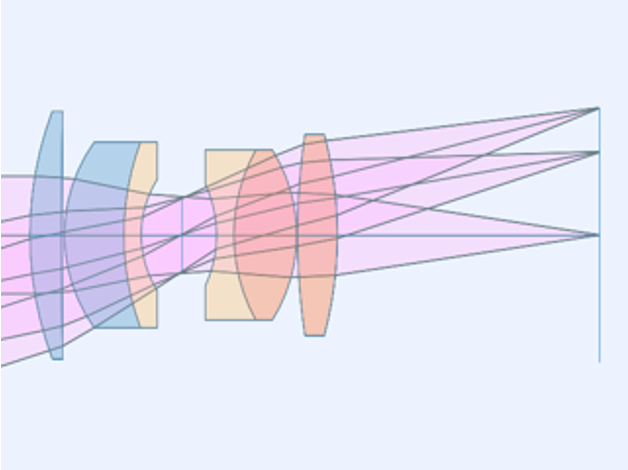}
    }
    \put(0,58){ 
    \small\textbf{(a)}
    }
    \put(100,58){ 
    \small\textbf{(b)}
  }
  \put(0,-2){ 
    \small\textbf{(c)}
  }
  \end{overpic}
  \end{flushleft}
  
  \renewcommand{\arraystretch}{1.2}
  \setlength{\tabcolsep}{4pt}  
  \resizebox{\linewidth}{!}{%
\scriptsize
  \begin{tabular}{|c|cc|cc|cc|cc|}
      \hline
   & \multicolumn{2}{c|}{\begin{tabular}{@{}c@{}} \textbf{Element 1} \\ \textbf{(L1)} \end{tabular}} &
     \multicolumn{2}{c|}{\begin{tabular}{@{}c@{}} \textbf{Element 2} \\ \textbf{(L2+L3)} \end{tabular}} &
     \multicolumn{2}{c|}{\begin{tabular}{@{}c@{}} \textbf{Element 3} \\ \textbf{(L4+L5)} \end{tabular}} &
     \multicolumn{2}{c|}{\begin{tabular}{@{}c@{}} \textbf{Element 4} \\ \textbf{(L6)} \end{tabular}} \\ \cline{2-9}
   & \textbf{Real} & \textbf{Pred.} & \textbf{Real} & \textbf{Pred.} & \textbf{Real} & \textbf{Pred.} & \textbf{Real} & \textbf{Pred.} \\ \hline
    $\Delta z$ [mm]      &  0.010 &  0.007 &  0.129 &  0.115 & $-0.895$ & $-0.897$ &  0.249 &  0.239 \\ 
    $\Delta x$ [mm]      &  0.931 &  0.885 &  0.471 &  0.539 &  0.634 &  0.588 &  0.135 &  0.082 \\ 
    $\Delta y$ [mm]      & $-0.527$ & $-0.525$ &  0.495 &  0.415 &  0.594 &  0.589 &  0.624 &  0.655 \\ 
    $\Delta\theta_x$ [deg] & $-0.588$ & $-0.596$ & $-0.602$ & $-0.613$ & $-0.419$ & $-0.409$ &  0.312 &  0.337 \\ 
    $\Delta\theta_y$ [deg] & $-0.711$ & $-0.707$ & $-0.192$ & $-0.192$ & $-0.513$ & $-0.497$ &  0.176 &  0.183 \\ \hline
  \end{tabular}
  }
  
  \caption{
    Results for the 6-lens system using Method~1. 
    (a) Training and validation MSE loss curves. 
    (b) Optical layout of the simulated six-lens photographic prime lens based on U.S. Patent US02194413-1. 
    (c) Predicted vs.\ ground-truth 5-DOF misalignments for one test sample (table).
  }
  \label{fig:spot_results}
\end{figure}

\textbf{Spot diagram model.}  
Method~1 accurately learns the relationship between spot pattern deformation and element-wise misalignments in the 6-lens system. As shown in Fig.~\ref{fig:spot_results}(a), both training and validation loss converge below an MSE of 0.005 in 500 epochs. The sharp change in slope around epoch 350 corresponds to the Reduce‑on‑Plateau scheduler reducing the learning rate by a factor of 0.1. The final model achieves a mean absolute error of 0.0317~mm in translation and 0.011$^\circ$ in tilt on the test set, evaluated on the physical (i.e., original-scale, unnormalized) misalignment values. Figure~\ref{fig:spot_results}(c) illustrates the predicted vs.\ ground-truth 5-DOF deviations for each of the four optical elements.  This level of accuracy would be challenging to achieve with analytical or manual optimization approaches, particularly for high-dimensional fault spaces.

\begin{figure}[htbp]

\begin{overpic}[width=0.74\linewidth]{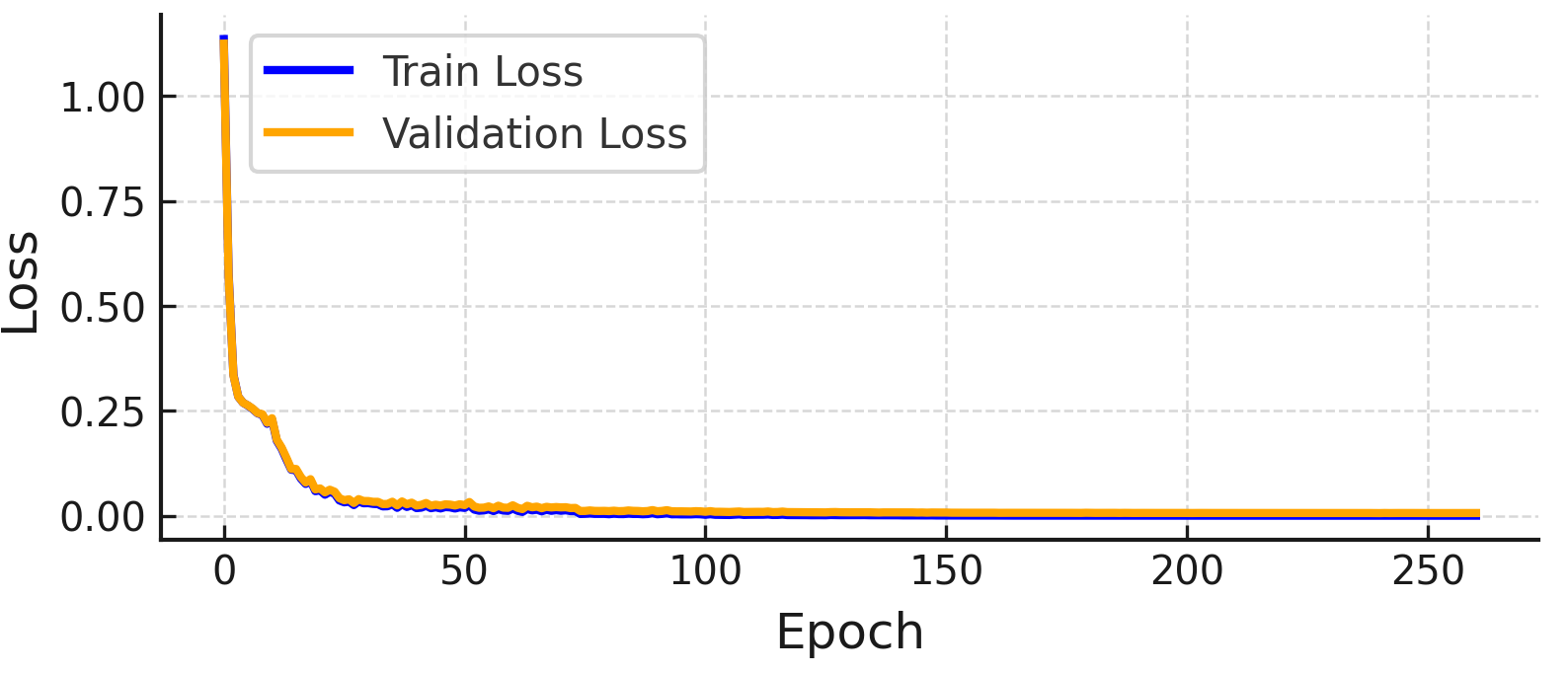}
  \put(99,-30){ 
    \includegraphics[width=0.26\linewidth]{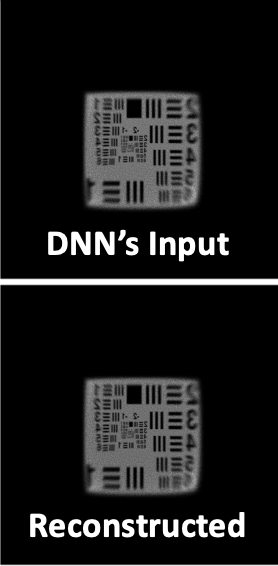}
  }
  \put(0,-30){ 
    \includegraphics[width=0.72\linewidth]{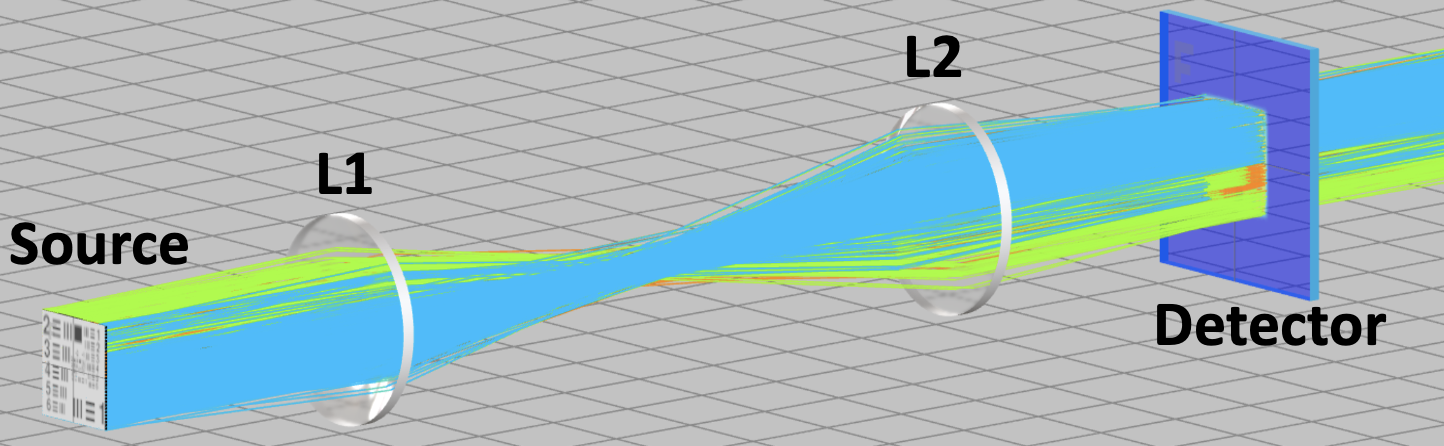}
  }
  \put(0,40){\small \textbf{(a)}}
  \put(0,1){\small \textbf{(b)}}
  \put(0,-33){\small \textbf{(c)}}
  \put(94,40){\small \textbf{(d)}} 
\end{overpic}

\vspace{9.5em}

\renewcommand{\arraystretch}{1.2} 
\setlength{\tabcolsep}{16pt}       
\resizebox{\linewidth}{!}{%
\scriptsize

\begin{tabular}{|c|cc|cc|}
  \hline
   & \multicolumn{2}{c|}{\textbf{Element 1 (L1)}} &
     \multicolumn{2}{c|}{\textbf{Element 2 (L2)}} \\ \cline{2-5}
   & \textbf{Real} & \textbf{Pred.} & \textbf{Real} & \textbf{Pred.} \\ \hline
  $\Delta x$ [mm]         & 1.840 & 1.879 & $-1.498$ & $-1.535$ \\ 
  $\Delta y$ [mm]         & $-1.822$ & $-1.844$ & 1.244 & 1.270 \\ 
  $\Delta\theta_x$ [deg]  & 2.703 & 2.761 & 1.048 & 1.024 \\ 
  $\Delta\theta_y$ [deg]  & 0.448 & 0.498 & $-2.289$ & $-2.258$ \\ \hline
\end{tabular}
}

\caption{
    Results for the 2-lens system. 
    (a) Training and validation MSE loss curves.  
    (b) Optical system schematic.  
    (c) Real vs.\ predicted 4-DOF misalignments for one test sample.  
    (d) Reconstructed image after simulating the system with the predicted misalignment vector (Reconstructed) and with the ground-truth misalignment vector (DNN's Input).
}
\label{fig:image2_results}
\end{figure}
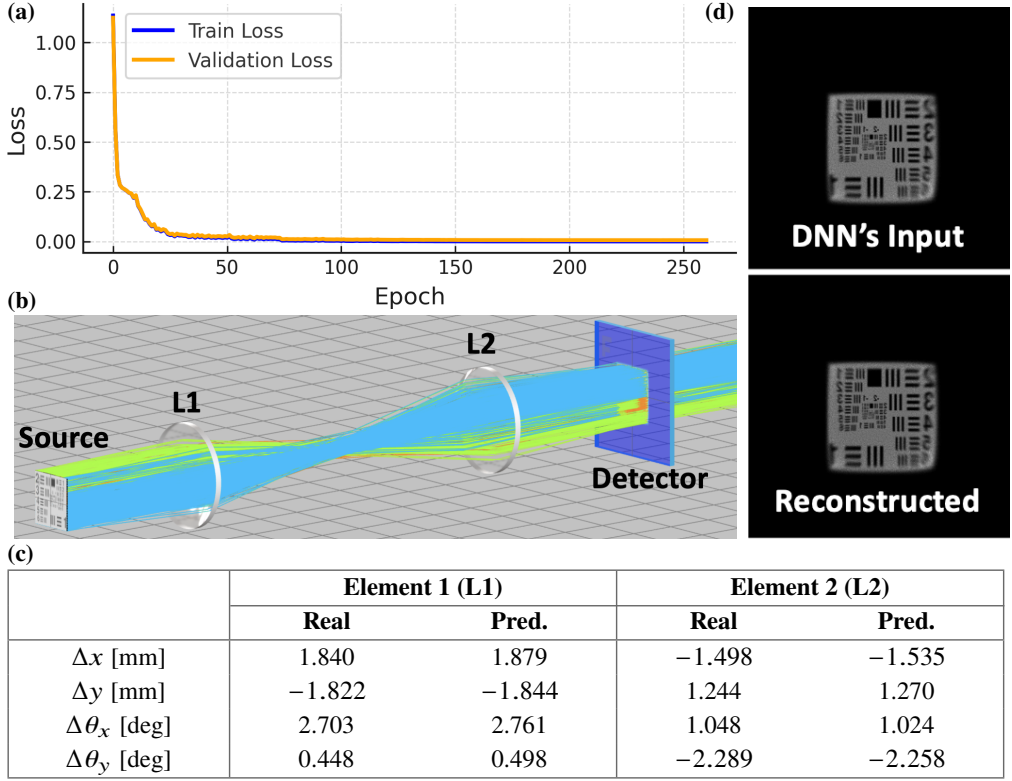
\textbf{Image-based model (2-lens system).}
The model trained on irradiance images from the 2-lens system achieves fast convergence with minimal overfitting. As shown in Fig.~\ref{fig:image2_results}(a), validation loss plateaus near 0.007 within a few dozen epochs. The model achieves a translational MAE of 0.044~mm and a rotational MAE of 0.121$^\circ$. Figure~\ref{fig:image2_results}(c) compares predicted and ground-truth misalignment parameters for a representative sample, showing close agreement across all four degrees of freedom. A qualitative comparison in Fig.~\ref{fig:image2_results}(d) shows that the reconstructed image, generated by simulating the system with the predicted misalignment vector, closely resembles the one produced using ground-truth parameters, demonstrating the model’s ability to recover precise alignment from raw irradiance data.

\begin{figure}[htbp]

\begin{overpic}[width=0.73\linewidth]{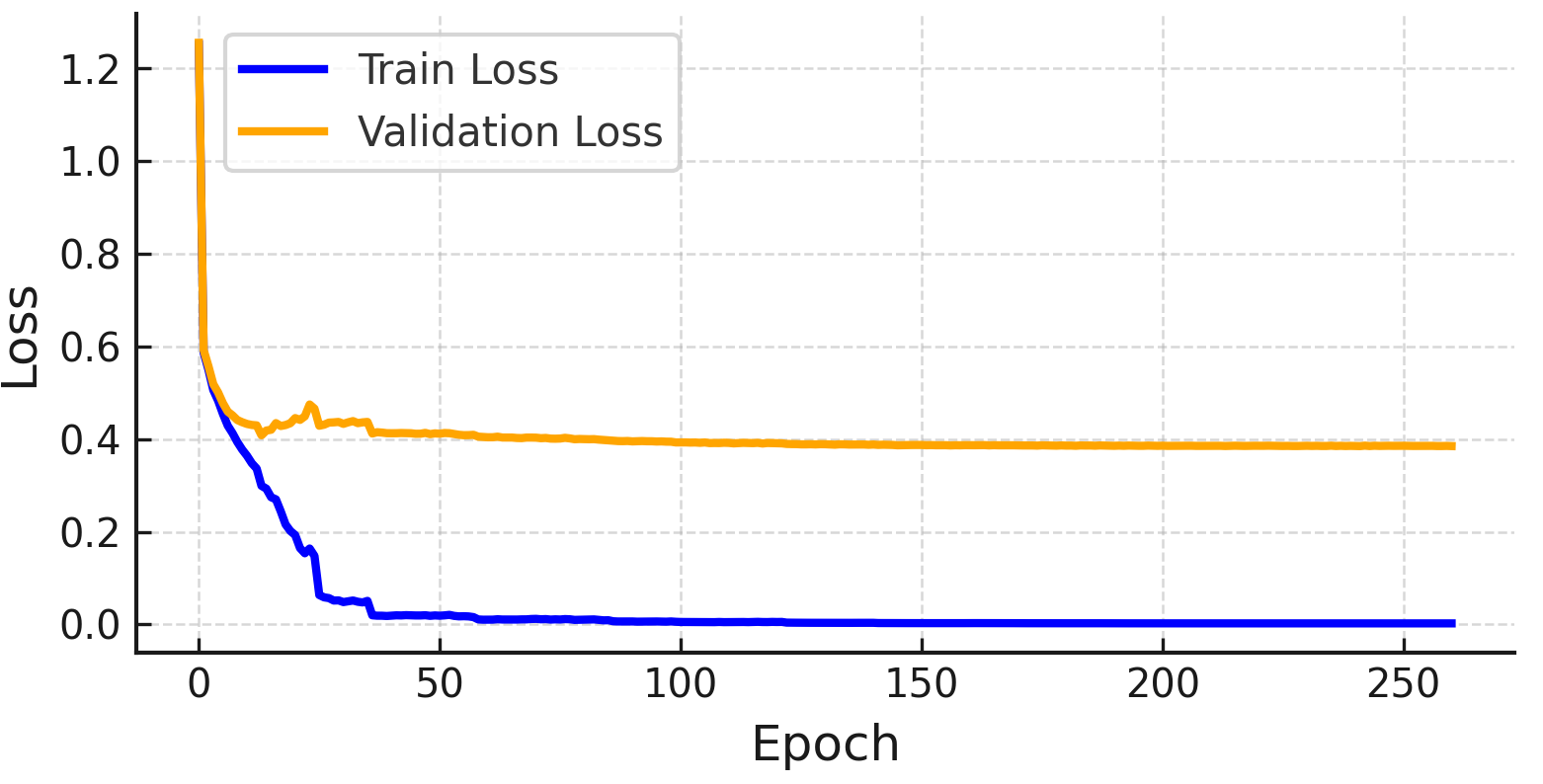}
  \put(99,-22){ 
    \includegraphics[width=0.27\linewidth]{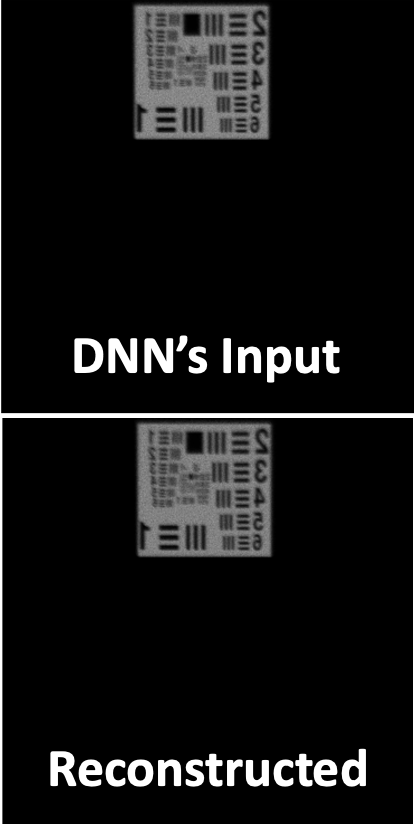}
  }
  \put(0,-22){ 
    \includegraphics[width=0.72\linewidth]{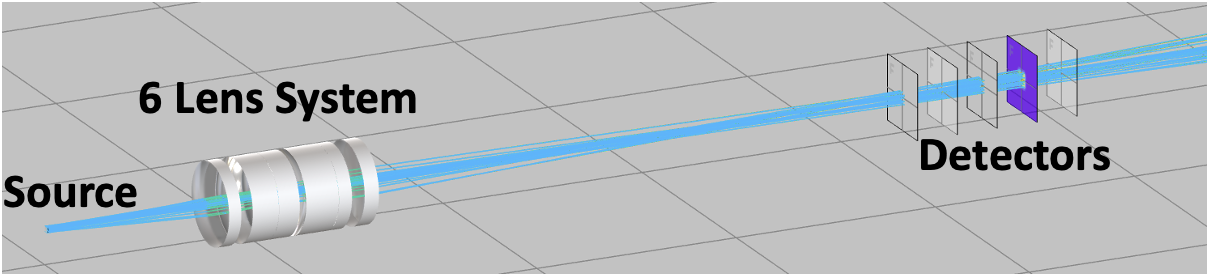}
  }
  \put(0,48){\small \textbf{(a)}}
  \put(0,1){\small \textbf{(b)}}
  \put(0,-25){\small \textbf{(c)}}
  \put(94,48){\small \textbf{(d)}} 
\end{overpic}

\vspace{7.2em}

\renewcommand{\arraystretch}{1.2} 
\setlength{\tabcolsep}{4pt}       

\resizebox{\linewidth}{!}{%
\scriptsize
\begin{tabular}{|c|cc|cc|cc|cc|}
  \hline
   & \multicolumn{2}{c|}{\begin{tabular}{@{}c@{}} \textbf{Element 1} \\ \textbf{(L1)} \end{tabular}} &
     \multicolumn{2}{c|}{\begin{tabular}{@{}c@{}} \textbf{Element 2} \\ \textbf{(L2+L3)} \end{tabular}} &
     \multicolumn{2}{c|}{\begin{tabular}{@{}c@{}} \textbf{Element 3} \\ \textbf{(L4+L5)} \end{tabular}} &
     \multicolumn{2}{c|}{\begin{tabular}{@{}c@{}} \textbf{Element 4} \\ \textbf{(L6)} \end{tabular}} \\ \cline{2-9}
   & \textbf{Real} & \textbf{Pred.} & \textbf{Real} & \textbf{Pred.} & \textbf{Real} & \textbf{Pred.} & \textbf{Real} & \textbf{Pred.} \\ \hline
  $\Delta x$ [mm]        & 0.151 & 0.143 & 0.126 & 0.049 & 0.084 & $-0.044$ & $-0.051$ & $-0.069$ \\ 
  $\Delta y$ [mm]        & $-0.232$ & $-0.133$ & 0.215 & 0.022 & $-0.153$ & 0.014 & $-0.039$ & $-0.088$ \\ 
  $\Delta\theta_x$ [deg] & 1.057 & 1.387 & $-1.089$ & $-1.881$ & 2.123 & 1.732 & 1.301 & 1.474 \\ 
  $\Delta\theta_y$ [deg] & $-1.949$ & $-2.134$ & $-1.406$ & $-1.421$ & $-2.355$ & $-2.232$ & 0.276 & 0.174 \\ \hline
\end{tabular}
}

\caption{
    Results for the 6-lens system. 
    (a) Training and validation MSE loss curves.  
    (b) Optical system schematic.  
    (c) Real vs.\ predicted 4-DOF misalignments for one test sample.  
    (d) Reconstructed image after simulating the system with the predicted misalignment vector (Reconstructed) and with the ground-truth misalignment vector (DNN's Input).
}
\label{fig:image6_results}
\end{figure}
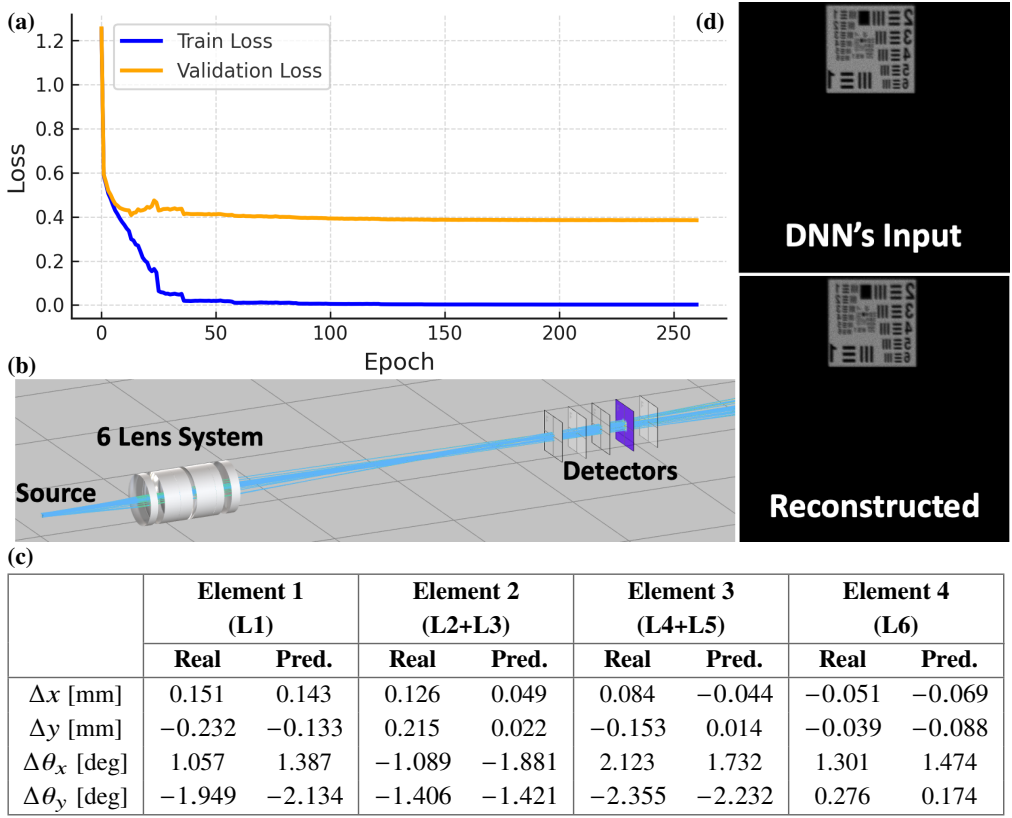

\textbf{Image-based model (6-lens system).}  
For the more complex 6-lens configuration, the model converges with higher final error due to increased complexity and data constraints. As shown in Fig.~\ref{fig:image6_results}(a), the validation loss stabilizes around 0.387 after 160 epochs, while the training loss continues to decrease. The changes in slope around epochs 30 and 60 in the training curve were caused by learning-rate reductions triggered by the Reduce-on-Plateau scheduler. Nevertheless, the model achieves a translational MAE of 0.089~mm and a rotational MAE of 0.505$^\circ$. 
Compared to the 2-lens system, prediction accuracy is reduced, as expected due to increased alignment complexity and the larger number of interacting elements. Still, Fig.~\ref{fig:image6_results}(c) shows that the model captures key trends in misalignment, with reasonable correspondence across both decenter and tilt parameters. The predicted correction improves the optical output, as seen in Fig.~\ref{fig:image6_results}(d), where the corrected irradiance image closely approaches the ideal aligned reference. 
While mild overfitting is observed due to limited data, the results confirm that the architecture generalizes well to multi-element alignment regression directly from sensor data.

\section{Summary}

We presented two complementary learning-based methods for predicting per-element misalignments in complex lens assemblies using only externally observable optical data. The first method relies on ray-traced spot diagrams and a fully connected network to recover 5-DOF alignment states with high accuracy across a six-lens system. The second method uses simulated irradiance images from 3DOptix and a hybrid ResNet–MLP architecture to estimate 4-DOF misalignments from physically realistic sensor views.

Both approaches demonstrate high accuracy using only synthetic, physics-based data. The spot diagram method provides fast convergence and strong precision, while the image-based pipeline generalizes well to practical sensor data and multiple optical systems.
These results show that deep networks can effectively learn the inverse mapping from optical measurements to alignment states in high-dimensional, nonlinear spaces, without requiring symmetry assumptions, internal access, or specialized metrology.

Crucially, the success of this approach is highly dependent on the informativeness of the simulated measurements. Parameters such as the placement of screens, range of rays, and location of the light source play a fundamental role in shaping the observability of lens misalignments. Poorly chosen configurations can degrade performance or render the inverse problem ill-posed. This sensitivity emphasizes the importance of co-designing data acquisition geometry alongside model training.

Finally, our approach can be readily applied to real-world systems, as the image-based method requires only a few standard grayscale images captured at different focus positions, easily acquired by any industrial camera within seconds. Because our simulations incorporate realistic physical effects, including surface scattering, polarization, and reflection, the trained model generalizes well to experimental conditions. A simple, one-time pixel-to-metric calibration enables hands-free alignment, making the method practical for both production and maintenance environments.

Compared to prior work, our method handles full multi-element perturbations and scales to realistic lens designs. These advances point to a scalable, simulation-driven paradigm for optical alignment that could enable faster, automated diagnostics in manufacturing and field settings. 

By replacing slow, manual methods with rapid, accurate, and automated diagnostics, this framework could greatly streamline industrial-scale optical assembly, substantially reducing production time, cost, and complexity across aerospace, biomedical, and consumer markets. More broadly, it stands as part of the sim-to-real revolution, turning designs perfected in high-fidelity simulation into plug-and-play alignment tools on the production floor.

\begin{backmatter}

\bmsection{Funding}
This work was supported by the AI and Data Science Center (TAD), Tel Aviv University.
\bmsection{Acknowledgment}
We acknowledge the 3DOptix team for technical support.

\bmsection{Disclosures}
The authors declare no conflicts of interest.

\bmsection{Data Availability}
All code and example datasets are available at:
\url{https://github.com/Tomerslortau/deep-learning-for-optics}.
\end{backmatter}

\clearpage

\appendix
\begin{center}
  {\LARGE\bfseries Supplementary Material}
\end{center}
\vspace{1em}
\addcontentsline{toc}{section}{Supplementary Material}
\renewcommand{\thefigure}{S\arabic{figure}}
\renewcommand{\thetable}{S\arabic{table}}
\renewcommand{\theequation}{S\arabic{equation}}
\renewcommand{\thesubsection}{S.\arabic{subsection}}

\section*{Optical System Configurations}

This section provides the detailed physical specifications for the simulated optical systems used in both the spot-diagram and image-based misalignment prediction methods.

\subsection{6-Lens System (Spot Diagram Method)}
\label{app:spot}
The 6-lens configuration is based on the photographic prime lens described in U.S. Patent \texttt{US02194413-1}. It consists of two cemented doublets and two singlets. The prescription is:

\begin{itemize}
  \item \textbf{Surface sequence:} \texttt{cgtctcgtcgtctcgtcgtctcgtct}
  \item \textbf{Surface curvatures (1/mm):}  
    [0.0437, 0.0020, 0.0970, 0.0580, 0.1480, $-$0.1290, 0.0880, $-$0.0960, 0.0230, $-$0.0390]
  \item \textbf{Axial spacings between surfaces (mm):}  
    [2.063, 0.165, 3.841, 1.113, 4.877, 1.113, 4.056, 0.051, 2.578, 16.892]
  \item \textbf{Refractive indices (550 nm):} [1.626, 1.626, 1.620, 1.654, 1.647, 1.647]
  \item \textbf{Semi-diameters (mm):}  
    [7.472, 7.189, 5.994, 4.417, 3.466, 3.100, 4.282, 5.139, 5.848, 6.093]
\end{itemize}

\begin{table}[ht]
  \centering
  \caption{6-lens spot-diagram system specifications.}
  \begin{tabular}{|c|c|c|c|c|}
    \hline
    Element & Type & Thickness & Index & Position \\
    \hline
    1 & Singlet (Surfaces 1–2) & 2.063 mm & 1.626 & 0 \\
    2 & Doublet (Surfaces 3–4) & 3.841 + 1.113 mm (cemented) & 1.620 / 1.654 & +0.165 \\
    3 & Doublet (Surfaces 5–6) & 1.113 + 4.056 mm (cemented) & 1.647 & +4.877 \\
    4 & Singlet (Surfaces 7–8) & 2.578 mm & 1.647 & +0.051 \\
    \hline
  \end{tabular}
  \label{tab:S1_element_specs}
\end{table}

The light source is 30 mm in front of the first surface. Rays (400 per field point, at 450 nm and 550 nm) are emitted from a 2×2 grid on the object plane. Two detectors, 16.892 mm and 26.892 mm behind the final surface, record the spots. Misalignments per element are:  
\(\Delta x,\Delta y\in[-1,1]\) mm; \(\theta_x,\theta_y\in[-1^\circ,1^\circ]\); and element-specific \(\Delta z\) bounds to avoid overlap.  

\subsection{2-Lens System (Image-Based Method)}
\label{app:image2}

This system uses two Edmund 50 mm-diameter singlets (plano-concave + plano-convex, 10 mm thick each) in 3DOptix, with a structured light source 99.99 mm before the first lens.

\begin{itemize}
  \item \textbf{Light source:} 465 nm; 25 mm×25 mm aperture; 3,333,333 rays (2D Gaussian, σ=1); 1951 USAF test-chart mask (792×612 px).
  \item \textbf{Lenses:}
  \begin{table}[H]
    \centering
    \begin{tabular}{|c|c|c|c|c|c|}
      \hline
      \textbf{Lens} & \textbf{Type} & \textbf{Radii (mm)} 
         & \textbf{Index} & \textbf{Position (mm)} & \textbf{Ref.\ Surface} \\
      \hline
      1 & Plano-Concave & Front: ∞; Back: 51.680 & 1.524 & 0 & Back \\
      2 & Plano-Convex  & Front: ∞; Back: –51.680 & 1.524 & +185.00 & Front \\
      \hline
    \end{tabular}
  \end{table}
  \item \textbf{Detector:} 60 mm×60 mm; 1000×1000 px; 96.13 mm behind last lens.
\end{itemize}

Misalignments: \(\Delta x,\Delta y\in[-2,2]\) mm; \(\theta_x,\theta_y\in[-3^\circ,3^\circ]\); no \(\Delta z\).

\subsection{6-Lens System (Image-Based Method)}
\label{app:image6}

This 6-lens assembly (BBAR-coated, two cemented doublets + two singlets) is from U.S. Patent \texttt{US02194413-1}, in 3DOptix with a 0.6 mm×0.6 mm rectangular source 20.095 mm before the first surface.

\begin{itemize}
  \item \textbf{Light source:} 465 nm; 3,333,333 incoherent rays (2D Gaussian, σ=1); 1951 USAF mask (792×612 px).
  \item \textbf{Lens specs:}
  \begin{table}[ht]
    \centering
    \caption{6-lens imaging system specifications.}
    \begin{tabular}{|c|c|c|c|c|c|c|}
      \hline
      Lens & Type & Thickness & Radii (mm) & Index & Diameter & Position \\
      \hline
      1 & Convex–Concave & 2.062 & F: 22.893; B: 636.592 & 1.633 & 10 mm & 0 \\
      2 & Convex–Concave & 3.841 & F: 10.323; B: 17.352  & 1.633 & 10 mm & +0.165 \\
      3 & Convex–Concave & 1.113 & F: 17.352; B: 6.749   & 1.628 & 10 mm & +0.020 \\
      4 & Concave–Concave & 1.113 & F: –7.739; B: 11.430 & 1.666 & 10 mm & +4.857 \\
      5 & Convex–Convex  & 4.056 & F: 11.430; B: –10.391& 1.655 & 10 mm & +0.020 \\
      6 & Convex–Convex  & 2.578 & F: 44.351; B: –25.425& 1.655 & 10 mm & +0.031 \\
      \hline
    \end{tabular}
    \label{tab:images_6lens_specs}
  \end{table}
  \item \textbf{Detectors:} Five screens (8 mm×8 mm, 1000×1000 px) at 67.5, 72.5, 77.5, 82.5, 87.5 mm from the last surface.
\end{itemize}

Misalignments: \(\Delta x,\Delta y\in[-0.25,0.25]\) mm; \(\theta_x,\theta_y\in[-3^\circ,3^\circ]\). Cropping thresholds vary per screen (10 W/cm² for screens 1–3, 9 W/cm² for 4, 5W/cm² for 5).

\subsection*{S.4 Additional Reconstructed Examples}
\label{app:recon}

\begin{figure*}[!htbp]
  \centering
  \includegraphics[width=0.8\textwidth]{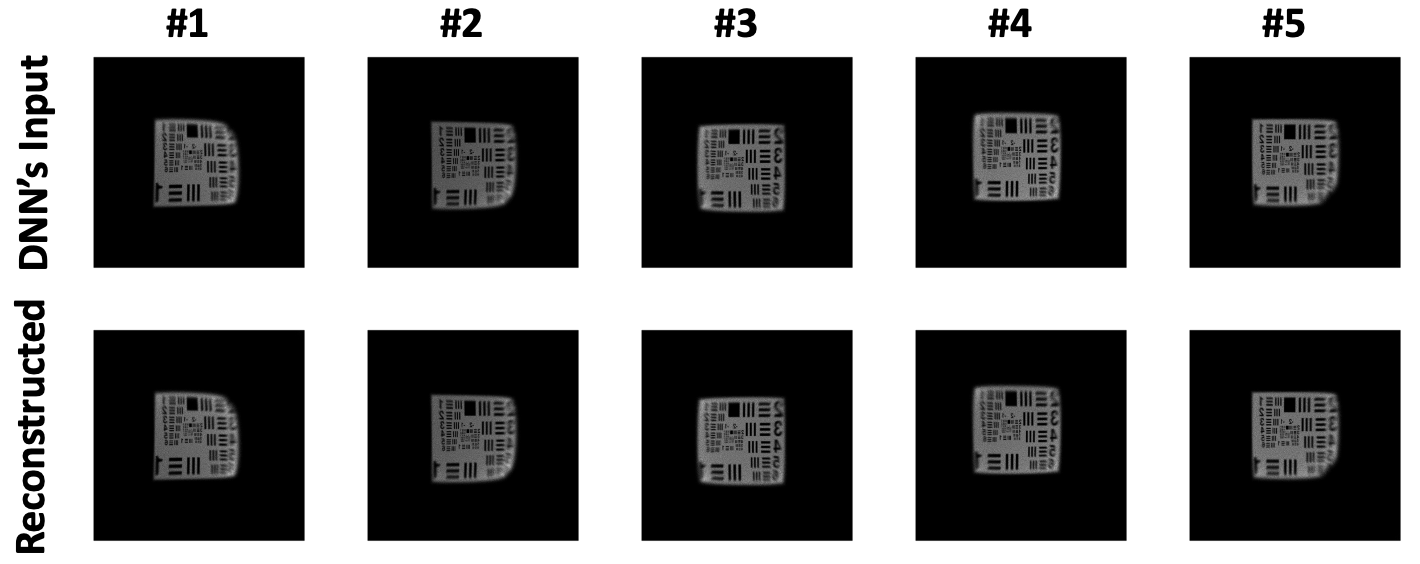}
  
  \vspace{0.5em}

\renewcommand{\arraystretch}{1.2}
\setlength{\tabcolsep}{3pt}
\scriptsize
\resizebox{\textwidth}{!}{%
\begin{tabular}{|c|c|cccc|cccc|}
  \hline
  \textbf{Type} & \textbf{\#} &
  \multicolumn{4}{c|}{\textbf{Element 1 (L1)}} &
  \multicolumn{4}{c|}{\textbf{Element 2 (L2)}} \\
  \cline{3-10}
  & & $\Delta x$ [mm] & $\Delta y$ [mm] & $\Delta\theta_x$ [deg] & $\Delta\theta_y$ [deg]
    & $\Delta x$ [mm] & $\Delta y$ [mm] & $\Delta\theta_x$ [deg] & $\Delta\theta_y$ [deg] \\
  \hline
  \multirow{5}{*}{\textbf{Real}} 
  & 1 & 1.840 & -1.822 & 2.703 & 0.448 & -1.498 & 1.244 & 1.048 & -2.289 \\
  & 2 & 1.986 & 0.842 & -0.266 & -0.022 & -0.676 & 0.661 & 2.418 & -2.514 \\
  & 3 & -0.910 & 0.514 & 0.505 & 2.126 & -1.201 & 1.164 & 0.829 & 1.208 \\
  & 4 & -0.308 & -0.774 & -2.511 & -1.011 & -0.398 & -0.149 & 2.573 & -2.297 \\
  & 5 & 1.900 & 1.718 & 0.085 & 2.675 & -1.699 & -1.311 & -1.838 & 2.117 \\
  \hline
  \multirow{5}{*}{\textbf{Predicted}} 
  & 1 & 1.879 & -1.844 & 2.761 & 0.498 & -1.535 & 1.270 & 1.024 & -2.258 \\
  & 2 & 1.975 & 0.826 & -0.172 & 0.053 & -0.677 & 0.689 & 2.382 & -2.566 \\
  & 3 & -0.984 & 0.534 & 0.501 & 2.276 & -1.131 & 1.164 & 0.808 & 0.892 \\
  & 4 & -0.198 & -0.759 & -2.532 & -1.187 & -0.499 & -0.182 & 2.500 & -1.942 \\
  & 5 & 1.793 & 1.736 & 0.037 & 2.809 & -1.630 & -1.339 & -1.948 & 1.879 \\
  \hline
\end{tabular}%
}

  \caption{2-lens system: Five examples of reconstructed images generated with predicted vs.\ ground-truth misalignments, alongside detailed numerical comparison.}
  \label{fig:recon_2lens}
\end{figure*}

\begin{figure*}[!htbp]
  \centering
  \includegraphics[width=0.8\textwidth]{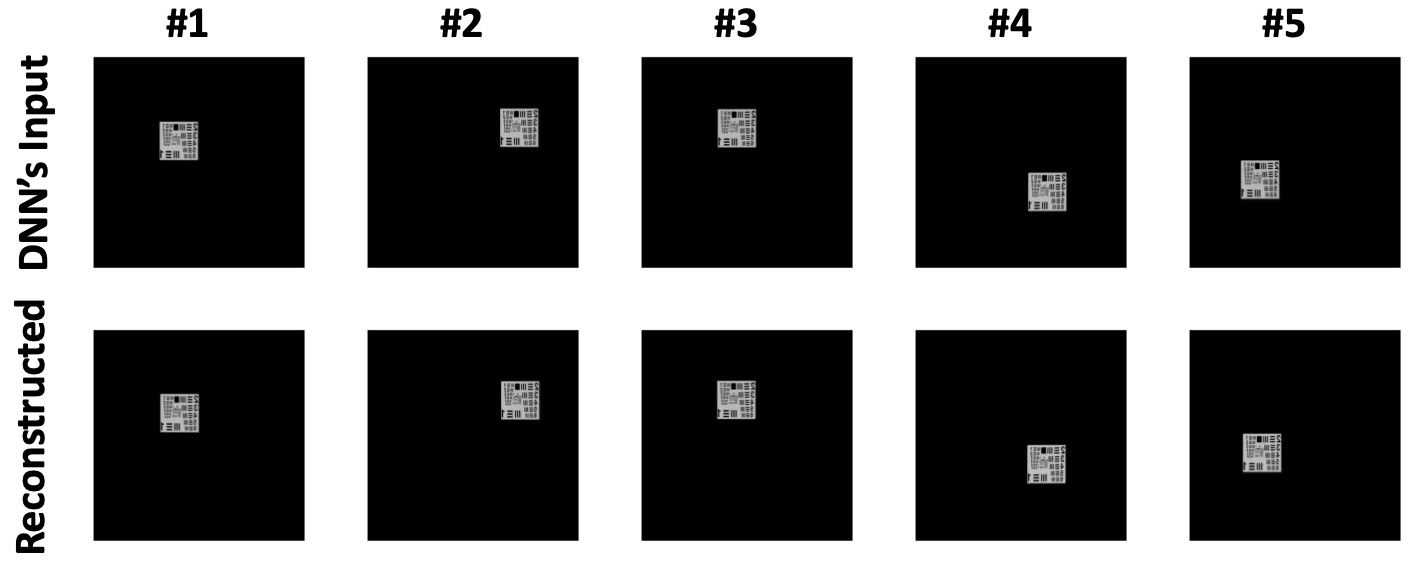}
\vspace{0.5em}

\renewcommand{\arraystretch}{1.2}
\setlength{\tabcolsep}{2pt}
\scriptsize
\resizebox{\textwidth}{!}{%
\begin{tabular}{|c|c|cccc|cccc|cccc|cccc|}
  \hline
  \textbf{Type} & \textbf{\#} &
  \multicolumn{4}{c|}{\textbf{Element 1 (L1)}} &
  \multicolumn{4}{c|}{\textbf{Element 2 (L2+L3)}} &
  \multicolumn{4}{c|}{\textbf{Element 3 (L4+L5)}} &
  \multicolumn{4}{c|}{\textbf{Element 4 (L6)}} \\
  \cline{3-18}
  & & 
  $\Delta x$ & $\Delta y$ & $\Delta\theta_x$ & $\Delta\theta_y$ &
  $\Delta x$ & $\Delta y$ & $\Delta\theta_x$ & $\Delta\theta_y$ &
  $\Delta x$ & $\Delta y$ & $\Delta\theta_x$ & $\Delta\theta_y$ &
  $\Delta x$ & $\Delta y$ & $\Delta\theta_x$ & $\Delta\theta_y$ \\
  & &
  {[mm]} & {[mm]} & {[deg]} & {[deg]} &
  {[mm]} & {[mm]} & {[deg]} & {[deg]} &
  {[mm]} & {[mm]} & {[deg]} & {[deg]} &
  {[mm]} & {[mm]} & {[deg]} & {[deg]} \\
  \hline
  \multirow{5}{*}{\textbf{Real}} 
  & 1 & 0.151 & -0.232 & 1.057 & -1.949 & 0.126 & 0.215 & -1.089 & -1.406 & 0.084 & -0.153 & 2.123 & -2.355 & -0.051 & -0.039 & 1.301 & 0.276 \\
  & 2 & -0.072 & 0.171 & -1.124 & -0.408 & 0.145 & -0.157 & -0.117 & 0.150 & 0.188 & -0.062 & -1.173 & -0.490 & -0.184 & 0.024 & 2.731 & 1.569 \\
  & 3 & -0.014 & -0.072 & -2.390 & -0.362 & 0.248 & -0.052 & -1.390 & 1.514 & 0.126 & -0.065 & -1.671 & 0.757 & 0.128 & -0.139 & 0.099 & 2.696 \\
  & 4 & -0.053 & -0.102 & -2.068 & 1.151 & 0.083 & 0.132 & -2.475 & -2.871 & 0.197 & -0.220 & -1.707 & -1.092 & 0.057 & -0.168 & -1.241 & 1.451 \\
  & 5 & 0.078 & 0.112 & 0.774 & 1.140 & -0.125 & -0.088 & 0.299 & -1.397 & -0.067 & -0.220 & 0.232 & -1.130 & -0.015 & 0.218 & -2.014 & 2.689 \\
  \hline
  \multirow{5}{*}{\textbf{Predicted}} 
  & 1 & 0.143 & -0.133 & 1.387 & -2.134 & 0.049 & 0.022 & -1.881 & -1.421 & -0.044 & 0.014 & 1.732 & -2.232 & -0.069 & -0.088 & 1.474 & 0.174 \\
  & 2 & -0.049 & 0.102 & -1.981 & 0.932 & 0.040 & 0.002 & 0.427 & 1.377 & 0.081 & 0.052 & -1.517 & -1.501 & -0.046 & 0.030 & 2.565 & 1.578 \\
  & 3 & 0.017 & -0.085 & -2.267 & -0.534 & 0.028 & 0.002 & -1.085 & 2.435 & 0.010 & -0.005 & -1.931 & 0.380 & 0.125 & -0.167 & 0.060 & 2.471 \\
  & 4 & 0.049 & -0.033 & -1.817 & 1.702 & 0.090 & 0.154 & -2.540 & -2.734 & 0.068 & -0.108 & -1.808 & -1.128 & 0.021 & -0.176 & -1.186 & 1.610 \\
  & 5 & 0.032 & 0.102 & 0.989 & 1.168 & -0.073 & -0.080 & 0.655 & -1.428 & -0.034 & 0.025 & -1.609 & -1.263 & 0.020 & 0.065 & -2.185 & 2.477 \\
  \hline
\end{tabular}%
}

  \caption{6-lens system: Five examples of reconstructed images (predicted vs.\ true misalignments).}
  \label{fig:recon_6lens}
\end{figure*}

Figures~\ref{fig:recon_2lens} and~\ref{fig:recon_6lens} illustrate representative reconstructions. The top row is the input image with random misalignments; the middle row is the simulation with the network’s predicted misalignments. The close visual agreement demonstrates the model’s accuracy.

\FloatBarrier

\bibliography{sample}

\begin{thebibliography}{10}
\newcommand{\enquote}[1]{``#1''}

\bibitem{malacara2007}
D.~Malacara, \emph{Optical Shop Testing} (John Wiley \& Sons, Hoboken, NJ, 2007), 3rd ed. Wiley Series in Pure and Applied Optics.

\bibitem{smith2008MOE}
W.~J. Smith, \emph{Modern Optical Engineering: The Design of Optical Systems} (McGraw-Hill, New York, 2008), 4th ed.

\bibitem{primot2000extended}
J.~Primot, V.~Daru, and B.~Gu\'{e}noche, \enquote{Extended hartmann test based on the pseudoguiding property of a hartmann maskcompleted by a phase chessboard,} {\protect\JournalTitle{Applied Optics}} \textbf{39}, 571--580 (2000).

\bibitem{crawford2008}
S.~M. Crawford, M.~Wells, and H.~Gajjar, \enquote{The use of primary mirrors as hartmann masks for in situ alignment of segmented mirror telescopes,} in \emph{Proc. SPIE,}  vol. 7012 (2008), p. 70123P.

\bibitem{wyant2010}
J.~C. Wyant, \enquote{Developments in optical testing technology during the last decade,} in \emph{Latin America Optics and Photonics Conf. (LAOP),}  (Optical Society of America, 2010), p. TuH1.

\bibitem{luna1999}
E.~Luna, A.~Cordero, J.~Valdez, \emph{et~al.}, \enquote{Telescope alignment by out-of-focus stellar image analysis,} {\protect\JournalTitle{Publications of the Astronomical Society of the Pacific}} \textbf{111}, 104--110 (1999).

\bibitem{Wang1993NeuralAlignment}
A.~J. Decker, M.~J. Krasowski, and K.~E. Weiland, \enquote{Neural-network-directed alignment of optical systems using the laser-beam spatial filter as an example,} Tech. Rep. NASA TP-3372, NASA Lewis Research Center (1993).

\bibitem{Zuo2022DeepLearningOpticalMetrology}
C.~Zuo, J.~Qian, S.~Feng, \emph{et~al.}, \enquote{Deep learning in optical metrology: a review,} {\protect\JournalTitle{Light: Science \& Applications}} \textbf{11}, 39 (2022).

\bibitem{Cote2021}
G.~Côté, J.~Lalonde, and S.~Thibault, \enquote{Deep learning‐enabled framework for automatic lens‐design starting‐point generation,} {\protect\JournalTitle{Optics Express}} \textbf{29}, 3841--3854 (2021). Published: January 25 2021.

\bibitem{Hegde2019}
R.~S. Hegde, \enquote{Accelerating optics‐design optimizations with deep learning,} {\protect\JournalTitle{Optical Engineering}} \textbf{58}, 065103 (2019).

\bibitem{Holters2016}
M.~Holters, A.~Gatej, S.~Haag, \emph{et~al.}, \enquote{Approach for self‐optimising assembly of optical systems,} {\protect\JournalTitle{International Journal of Computer Integrated Manufacturing}} \textbf{29}, 1227--1237 (2016).

\bibitem{Min2024}
H.~Min, Y.~Son, and Y.~Choi, \enquote{Determining optimal assembly condition for lens module production by combining genetic algorithm and c-blstm,} {\protect\JournalTitle{Processes}} \textbf{12}, 452 (2024).

\bibitem{Gu2020}
Z.~Gu, Y.~Wang, and C.~Yan, \enquote{Optical system optimization method for as‐built performance based on nodal aberration theory,} {\protect\JournalTitle{Optics Express}} \textbf{28}, 7928--7942 (2020).

\bibitem{Liu2024ActiveAlignment}
H.~Liu, W.~Li, S.~Gao, \emph{et~al.}, \enquote{Application of deep learning in active alignment leads to high-efficiency and accurate camera lens assembly,} {\protect\JournalTitle{Optics Express}} \textbf{32}, 43834--43849 (2024).

\bibitem{Jia2021TelNet}
P.~Jia, X.~Wu, Z.~Li, \emph{et~al.}, \enquote{Point spread function estimation for wide-field small-aperture telescopes with deep neural networks and calibration data,} {\protect\JournalTitle{Monthly Notices of the Royal Astronomical Society}} \textbf{505}, 4717--4725 (2021).

\bibitem{Baslar2024MisalignmentNN}
I.~Ba\c{s}lar and M.~Dursun, \enquote{Improving neural-network-based prediction models for misalignment in off-axis three-mirror anastigmat telescopes,} {\protect\JournalTitle{Applied Optics}} \textbf{63}, 7747--7755 (2024).

\bibitem{Wu2022MLAlignment}
Z.~Wu, Y.~Zhang, R.~Tang, \emph{et~al.}, \enquote{Machine learning for improving stellar image-based alignment in wide-field telescopes,} {\protect\JournalTitle{Research in Astronomy and Astrophysics}} \textbf{22}, 015008 (2022).

\bibitem{Rothe2023SkySurvey}
M.~Zhang, P.~Jia, and Z.~Li, \enquote{Perception of misalignment states for sky survey telescopes with digital twin and deep neural networks,} {\protect\JournalTitle{Optical Engineering}} \textbf{62}, 44054 (2023).

\bibitem{Gungor2025ToleranceAware}
J.~Dai, L.~Chen, and T.~Xue, \enquote{Tolerance-aware deep optics,} {\protect\JournalTitle{arXiv preprint}} \textbf{arXiv:2502.04719} (2025).

\bibitem{Hashimoto:24}
R.~Hashimoto, S.~Matsuura, and Y.~Iida, \enquote{Method for optical adjustment with deep learning to quantitatively predict misalignment in optics,} {\protect\JournalTitle{Appl. Opt.}} \textbf{63}, 6794--6805 (2024).

\bibitem{RayOpt}
{RayOpt}, \enquote{Rayopt,} \url{https://github.com/quartiq/rayopt} (2020). GitHub repository.

\bibitem{3doptix2025}
{3DOptix Ltd.}, \enquote{3doptix: Cloud-based optical design and simulation platform,}  (2025).

\end{thebibliography}

\end{document}